# Profiling lung cancer patients using electronic health records

**Ernestina Menasalvas Ruiz**[1,2]
ernestina.menasalvas@upm.es

**Juan Manuel Tuñas**[1]
juan.tunas@ctb.upm.es

**Guzmán Bermejo**[1]
guzman.bermejo@ctb.upm.es

**Consuelo Gonzalo Martín**[1]
consuelo.gonzalo@upm.es

**Alejandro Rodríguez-González**[1,2]
alejandro.rg@upm.es

**Massimiliano Zanin**[1]
massimiliano.zanin@ctb.upm.es

**Cristina González de Pedro**[3]
cristinagdp@live.com

**Marta Méndez**[3]
mmendezo@salud.madrid.com

**Olga Zaretskaia**[3]
olga.zaretskaia@salud.madrid.com

**Jesús Rey**[4]
jrey@idiphim.com

**Consuelo Parejo**[4]
cparejo@idiphim.com

**Juan Luis Cruz Bermudez**[2,4]
jlcruz@idiphim.com

**Mariano Provencio**[3]
mprovencio@idiphim.com

[1]Universidad Politécnica de Madrid
Centro de Tecnología Biomédica
Campus de Montegancedo
Pozuelo de Alarcón, 28223, Madrid, Spain

[2]Universidad Politécnica de Madrid
Escuela Técnica Superior de Ingenieros Informáticos.
Departamento de Lenguajes y Sistemas Informáticos e Ingeniería del Software

Campus de Montegancedo
Boadilla del Monte, 28660, Madrid, Spain

[3]Hospital Universitario Puerta de Hierro de Majadahonda
Medical Oncology Service
Majadahonda, 28222, Madrid, Spain

[4]Hospital Universitario Puerta de Hierro de Majadahonda
Unidad TIC IIS
Majadahonda, 28222, Madrid, Spain

# ABSTRACT

If Electronic Health Records contain a large amount of information about the patient's condition and response to treatment, which can potentially revolutionize the clinical practice, such information is seldom considered due to the complexity of its extraction and analysis. We here report on a first integration of an NLP framework for the analysis of clinical records of lung cancer patients making use of a telephone assistance service of a major Spanish hospital. We specifically show how some relevant data, about patient demographics and health condition, can be extracted; and how some relevant analyses can be performed, aimed at improving the usefulness of the service. We thus demonstrate that the use of EHR texts, and their integration inside a data analysis framework, is technically feasible and worth of further study.

## 1. Introduction

Cancer is still a major problem from a health and economic perspective. According to Eurostat, cancer causes one out of every four deaths in Europe [1]. More concretely, lung cancer causes an estimated 20.8% of all cancer deaths, with more than 266.000 lives lost in Europe in 2011. Beyond such high mortality, lung cancer was the one with highest economic cost (€18,8 billion, 15% of overall cancer costs [2]) when including both the social costs derived from the productivity loss due to the anticipated death, and the direct health care costs. Still, the main cost driver is the interaction with healthcare professionals, such that hospitalizations alone account for a 47% of the total cost, even above oncology treatments and drugs themselves [3].

The Medical Oncology Department at Puerta de Hierro University Hospital (HUPHM) is addressing this problem through a reduction of unnecessary visits to ER and unscheduled appointments. This has been achieved by installing a Telehealthcare Phone Service (TPS), attended by oncology nurses, and located at the medical oncology clinic. This service avoided 47.4% of the patient visits to the ER, 46.8% of the unscheduled visits to Medical Oncology, and 5.5% of the visits to primary care doctors [4]. In spite of these positive results, a more detailed analysis of the profile of the calling patient is still required, including of his/her behavior and evolution, to further improve the service and to develop new means for interacting with the patient, such as mobile apps.

The aim of the current study is to describe the main characteristics of those lung cancer patients who have used the aforementioned HUPHM's assistance call service. A first analysis of the service has already been presented in [5], specifically by studying a limited set of characteristics about the calling patients, and extracting some descriptive and predictive insights on them. The limitedness of the information thereby contained was nevertheless the major barrier towards more useful results. In order to tackle this point, the present study focuses on extracting and integrating information from other complementary data sets, thus enlarging the scope of the subsequent analysis. We also focus on lung cancer patients, for being the most representative part of the whole population, and for providing a homogeneous cohort of patients.

The original TPS' information has been enriched through an analysis of the data contained in patient's Electronic Health Records (EHRs). While HUPHM has been using an EHR system since the end of 2008, this was not designed to be used in clinical research. Except for demographic data and those related to assistance activity, most of the information acquired by the physicians is recorded in a non-structured format (free text), which hinder their direct use for analysis. The hospital generates about 2,5 million clinical notes per year, and it has been estimated that lung cancer patients alone have generated more than 700,000 clinical notes in free text since 2008. On top of this, the hospital medical services also elaborate registries of assistance cases by using office software, which include structured clinical data that can be of interest for both clinical and research activities. In this context, the application of NLP tools to extract the information from the free text clinical notes can have an enormous potential.

With respect to [5], this work thus include an application of text mining techniques to enable the automatic extraction of a large number of features from EHRs written in Spanish. Thanks to this technology, the second objective here pursued is demonstrating that NLP can be integrated into a data mining pipeline, specifically to extract features enabling a deeper understanding of the patients making use of the TPS, and eventually improving the latter.

## 2. Data Description

The data that have been used in the study come from two different sources. The first one, the Calls Dataset (CD), is a subset of the one used in the previous work [5], which originally contained a total of 1242 call records registered from 568 distinct patients. From these, the chosen subset included only those patients that were diagnosed with lung cancer and performed a call to the service (a total of 64 patients).

The second source is the Electronic Health Record Dataset (EHRD). This dataset was generated by retrieving all the records (clinical notes and reports) corresponding to those 64 lung cancer patients. Clinical notes contain information about those services that have been visited by the patient at the hospital. They are always written by a professional (*e.g.* physicians, nurses, social services people, etc.). These include highly detailed information about the patients, as well as the different processes that have been performed, their results, and so forth. On the other hand, reports contain a more summarized information, which is usually generated when a specific clinical process has been completed. Figures 1 and 2 reports two examples of the type of structure and information that are typically present in clinical notes and reports provided by HUPHM.

```
<u><b>ENSAYO OAM9861g (PULMóN).- CICLO 3<br /></b></u><br
/><b>TRATAMIENTO:<br />ERLOTINIB 100 mg/día <br />+<br
/>METMAB/PLACEBO 15 mg/kg IV cada 3 semanas </b><br /><br />NOTA
ACLARATORIA: La paciente sólo ha recibido una línea de
quimioterapia previa a la inclusión en el ensayo
(CDDP-VINORELBINA).<br /><br />o PS:0<br />o Exploración
física: normal salvo rash.<br />o Adenopatías,
hepatomegalia, esplenomegalia: no palpables.<br /><br />o Talla:
1,66m<br />o Peso: 66,7Kg<br />o TA:110/77mmHg<br />o FC:106lpm<br />o
FR:14rpm<br />o Tª (axilar): 35,5ºC<br /><br
/><u>*Medicación concomitante:</u><br />Doxiciclina
100mg/24horas.<br />Peitel cada 12 horas.<br /><br
/><u>*Acontecimientos adversos:</u><br />Astenia leve.<br />Rash grado
2-3. Intensidad moderada-severa aunque no abarca más del 50%.
Afecta cara y más levemente la región superior del
tórax.<br /><br /><u>*Analítica:</u><br /><u>Incidencia:
Suero muy hemolizado<br /></u>BIOQUÍMICA: Glucosa 101.00 mg/dl (60.0 –
100.0), Urea 43.00 mg/dl (21.0 – 50.0), Creatinina 0.30 mg/dl (0.5 –
0.9), Calcio 8.00 mg/dl (8.7 – 10.3), Sodio 137.00 mmol/L (135.0 –
145.0), <u>Potasio 5.30 mmol/L</u> (3.5 – 5.0), Cloruro 106.00 mmol/L
(101.0 – 111.0), Bilirrubina total 0.40 mg/dl (0.3 – 1.1), ALT (GPT)
30.00 U/L (6.0 – 40.0), <b>AST (GOT ) 42.00 U/L</b> (6.0 – 40.0),
gamma-Glutamiltransferasa 24.00 U/L (6.0 – 36.0)<br
/>HEMOGRAMA: Leucocitos 6.58 x10E3/microL (4.0 – 11.5),
Neutrófilos 4.93 10E3/microL (1.5 – 7.5), Linfocitos 0.72
10E3/microL (1.2 – 4.0), Monocitos 0.75 10E3/microL (0.2 – 1.0),
Eosinófilos 0.11 10E3/microL (0.0 – 0.4), Basófilos 0.04
10E3/microL (0.0 – 0.2), Hematíes 5.28 10E6/microL (4.0 – 5.4),
Hemoglobina 15.00 g/dL (12.0 – 17.0), Hematocrito 43.20 % (41.0 –
53.0), V.C.M 81.80 fL (82.0 – 97.0), H.C.M 28.40 pg (26.0 – 31.0),
C.H.C.M 34.70 g/dL (32.0 – 36.0), Plaquetas 297.00 10E3/microL (150.0
– 400.0)<br /><br />Iniciamos Afatinib a 30mg.<br />Control
clínico analítico el 03/07/2019<br />
```

Fig. 1. Excerpt of the raw information contained in a clinical note, describing a treatment and corresponding clinical analyses.

```
BPIFECHA INGRESO: 07/03/2019 15:06
BPIFECHA ALTA: 14/03/2018
BPI
BPIMÉDICO RESPONSABLE INFORME: Dr. García (Médico Adjunto). Dra.
Pérez (Médico Residente)
BPI
BPI
Motivo de Ingreso:
Disnea y fiebre.

Antecedentes Personales
NO alergias medicamentosas conocidas.
- No HTA, no DM, no DL.
- Ex fumadora desde dice/2011 de 40 cigarrillos / día; exADVP desde
el 1999 de heroína y cocaína (VIH negativo).
- VHC con carga indetectable.
- IQx: muelas y quiste mandibular, amigdalectomía.
- Tratamiento habitual: ninguno.

Antecedentes Familiares:
- Padre fallecido por ca. pulmón con 46 años.

HISTORIA ONCOLOGICA:
Varón de 50 años diagnosticado de Adenocarcinoma de pulmón tras
hallar Rx de Tórax de preoperatorio de Cirugía bucal lesión
sospechosa. En un año había presentado varios episodios de
bronquitis que ha precisado tratar con antibioterapia.VTSe han
realizado hasta ahora las siguientes pruebas complementarias:
- Rx tórax (solo informe): opacidad mediastínica inferior derecha.VT
- Fibrobroncoscopia (23-05-11): Masa endobronquial en el bronquio
lobar inferior.VTInforme Anatomía Patológica:
ASPIRADO BRONQUIAL : Citología SOSPECHOSA para células malignas
BIOPSIA por broncoscopia BRONCOSCOPIA: Sin hallazgos
concluyentes. En la muestra remitida se incluyen framentos de
material vegetal
```

Fig. 2. Excerpt of the raw information contained in a clinical, with a synthetic overview of the history and status of the patient.

In both cases the information is written by the professional using free-text. The number of clinical notes that were processed for this study was 1160 (between 1 and 76 per patient, mean of 18, mode of 3) and the number of reports was 977 (between 1 and 60 per patient, mean of 15, mode of 4), for a total of 2137 medical texts processed.

### 3. Methodology

The present study has been organized into two main steps. The first one, as already described in the Data Description section, was based on the retrieval of the specific electronic health records of the subject population. Afterwards, the second step involved the execution of our Clinical Knowledge Extraction System (C-liKES).

C-liKES is a framework that has been developed on top of Apache UIMA, and which has been based on a legacy system named H2A [6]. The framework is a text-mining system that has been developed to ingest clinical information in free-text format; and to yield a structured output that can both support complex queries and be used for applying more in-depth analytics such as the application of machine learning techniques.

The general architecture of C-liKES is depicted in figure 3.

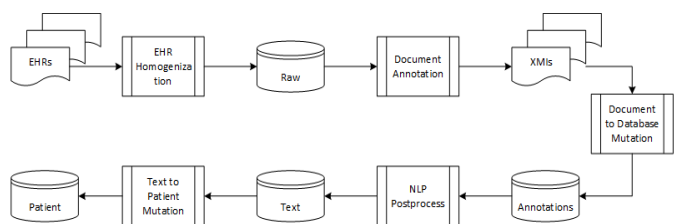

Fig. 3. General architecture of C-liKES.

The main elements of C-liKES and a brief description of their behavior and used data sources

are explained below:

1. **EHR Homogenization**: takes EHRs in different formats and makes the required transformations in order to normalize them and harmonize the structure. Afterwards it stores the information in a relational database.
2. **Document Annotation**: reads patient reports from the harmonized database and performs the NLP annotation processes. These include tokenization, part-of-speech tagging, chunking and named entity recognition. The output obtained from this stage is stored in structured XML Metadata Interchange (XMI) documents.
3. **Document to Database Mutation**: process in charge of transforming the XMI documents obtained in the previous phase to a database, in order to enable the execution of queries.
4. **NLP Postprocess**: several natural language processing components are executed, including negation detection and named entities context relationship discovery. The results are stored in a relational database that is organized mimicking the document internal structure.
5. **Text to Patient Mutation**: a semantic analysis is performed, yielding a database that is centered on the patient characteristics and their evolution though time.

As has been seen in the general architecture, C-liKES ingests textual information provided by the hospital (in this case HUPHM) and applies the corresponding transformation process that allows the execution of the rest of the pipeline.

Once the transformation process, which is the only module dependant on the data provider, is executed, the rest of the system starts the structuration of the clinical records. The output produced by the system is a local database containing the information retrieved. The system has been designed to extract the information surrounding the "Patient" entity, as the latter is naturally the main element in any clinical record. For a given patient it is possible to retrieve basic demographic information, diagnosis date, familiar and personal antecedents, occupation, comorbidities, toxic habits, previous treatments and event-time information. The event-time information refers to the information that is attached to a specific event that occurs at a specific time. To illustrate this latter concept, a clinical note that includes a psychological evaluation can be part of the clinical history of the patient; such evaluation is associated with a date, as it could have happened before or after the diagnosis date, this difference being of relevance from a clinical perspective. When available, additional information such as treatments or clinical entities (symptoms, diagnosis/diseases, diagnosis tests, results) is also structured and attached to the specific event.

At the end of the process, the framework stores the structured information in a relational database, which can be accessed through simple or complex queries.

## 4. Analysis and results

The data to perform the following analyses have been retrieved by the structured database populated by C-liKES from the Electronic Health Record Dataset (EHRD) to extend the Calls Dataset (CD). Some variables have directly been retrieved, as:

- Smokers (current and past) vs. no-smokers;

- Under treatment of tyrosine-kinase-inhibitor (TKI);
- City region - this variable has been obtained from the structured EHR and is represented by the ZIP code.

On the other hand, some of them required a more complex extraction procedure, like:

- Age of diagnosis;
- Eastern Cooperative Oncology Group scale of Performance Status – ECOG Performance Status or ECOG in short [7]: a scale describing the patient level of ability to care for themselves, daily activity, and physical ability.

In most of the cases, these two variables do not explicitly appear in the clinical notes, making it necessary to look forward for other indicators or related terms that allow an indirect estimation of their values.

All the previously described variables extend the ones that were already analyzed in the previous study [5]:

- Gender;
- Age at call time;
- Stage of the cancer.

With the help of the chosen variables, it is possible to describe and cluster the profiles of patients with lung cancer. Specifically, the age (both at diagnostic and at the time of the call), gender and city provide us with basic demographic data. These, together with information on the severity of the disease (stage) and the patient's performance status, allow us to describe the profiles according to the severity of the cancer, the general condition of the patient and the demographic information. This is expected to yield a much more detailed view on the disease, when compared with standard clinical profiles only leveraging on gender and age groups.

In what follows, the analysis of the resulting data set is organized in three phases: *i*) a statistical overview of the data, *ii*) an analysis of the relationship between TKI and smoking habit, and *iii*) a characterization of the profile of callers. Note that what here reported does not exhaust all possible analyses that can be performed on the data; on the contrary, this is meant as an example of the kind of knowledge that can be extracted thanks to the application of NLP to EHR.

A. **Statistical overview of the data**

As a first step, and in order to support the subsequent analyses, we proceeded to the development of a descriptive analysis of the population with lung cancer that used the oncology call service. The distribution of the main variables is shown in figures 4 to 9.

It is worth noting that these distributions cannot be analyzed alone, but that instead require a comparison against the full population, *i.e.* composed of callers and non-callers. Nevertheless, some general considerations can still be extracted, as done below; and this information is readily of interest for physicians and the hospital staff, as it provides them with an overview of the interactions between users and the system.

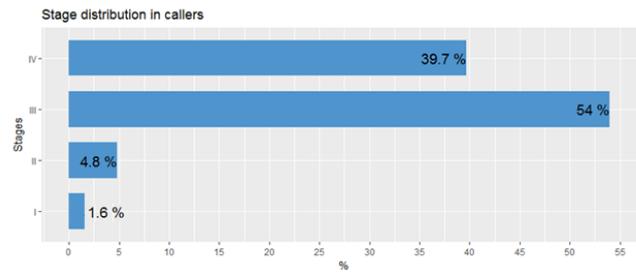

Fig. 4. Distribution of stages in callers.

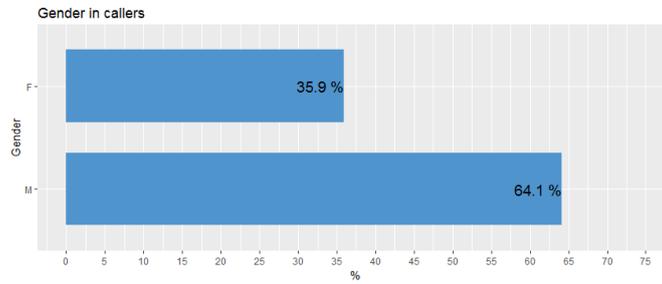

Fig. 5. Distribution of the gender of callers.

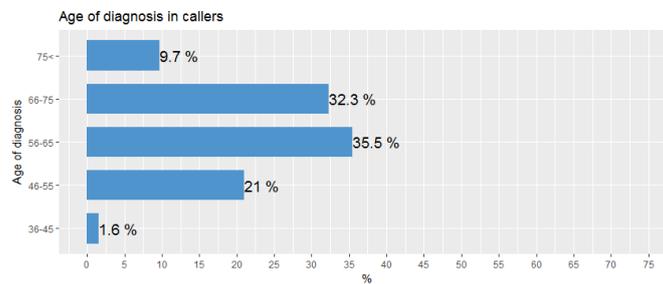

Fig. 6. Distribution of the age of callers.

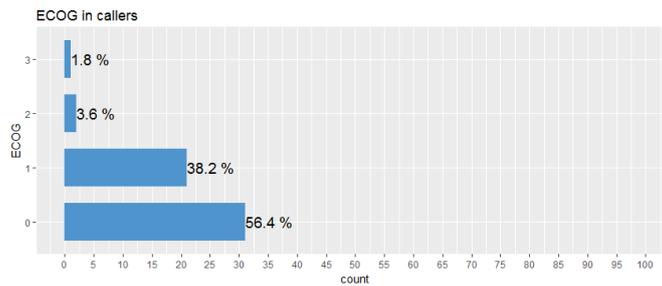

Fig. 7. Distribution of ECOG for callers.

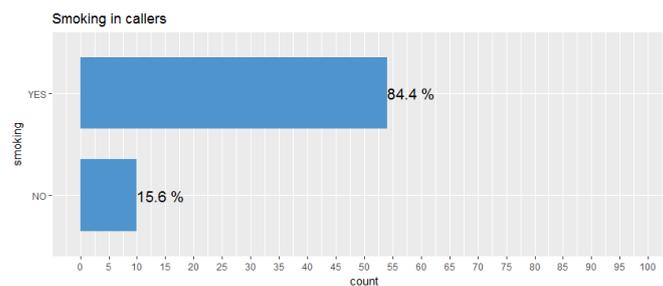

Fig. 8. Proportion of smokers and non-smokers.

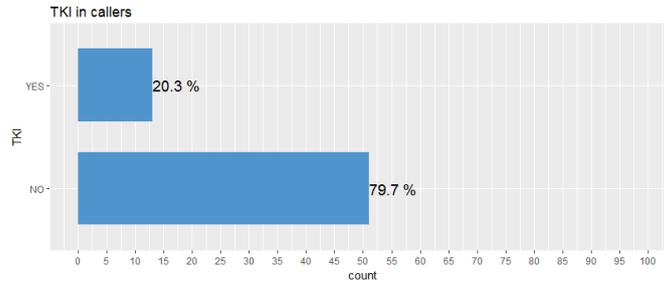
Fig. 9.  Proportion of patients treated with TKI.

B.  **Relationship between TKI and smoking habit**

As a second step, we here focus on the presence of a specific relationship between the TKI variable (*i.e.* if the patient is treated with TKI or not) and the smoking habit. This choice is supported by two considerations. First of all, there is an already proven and well-known cause-effect relationship between smoking and lung cancer [8], [9]. Second, it is also recognized that non-smoking patients with lung cancer share in their majority the characteristic of having a mutation in the genes controlling the epidermal growth factor receptor, thus making them good targets for tyrosine-kinase-inhibitor treatments [10].

|             |     | **TKI treatment** |     |
|-------------|-----|-------------------|-----|
|             |     | Yes               | No  |
| **Smoking** | Yes | 30                | 555 |
|             | No  | 25                | 88  |

Tab. 1. Distribution of patients according to smoking habit and TKI treatment.

Table 1 reports the contingency table for all patients, according to their smoking habit and the presence of a TKI treatment. The high number of patients smoking and not treated with TKI confirms the aforementioned relationship between these two aspects. From a statistical point of view, a Pearson's $\chi^2$ test yields a *p*-value of 8.3e-10, thus confirming the significance of results. One further obtains a Phi-Coefficient of 0.232, a Contingency Coefficient of 0.226, and a Cramer's V of 0.232. All these values corroborate the existence of an inverse relationship between both variables (non-smokers related with TKI treatment / smokers related to the absence of TKI treatment), in line with what present in the literature.

C.  **Profiling the callers**

We finally present a demographic analysis of the characteristics of callers, through a multivariate visualization of four of the previously described variables. First, figure 10 reports the distribution of the gender according to the severity of the stage in the two advanced stages (III and IV). Then, in figure 11, the variable 'age at diagnosis' is included in the analysis, to observe if there are differences in the gender distributions according to age for these two stages.

The interpretation of figure 11 must take into account the information represented in figure 10, and specifically the gender distribution by stages. For example, it seems that women are more likely to call in younger age in comparison with men, as the distribution in the younger age group shows more proportion of women than the 'all-ages' distribution in both stages. Furthermore, this value is also greater than the total percentage of women enrolled in the study.

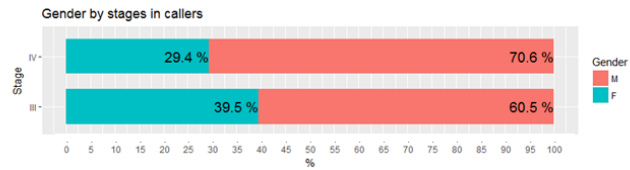

Fig. 10.  Distribution of gender by advanced stages in callers.

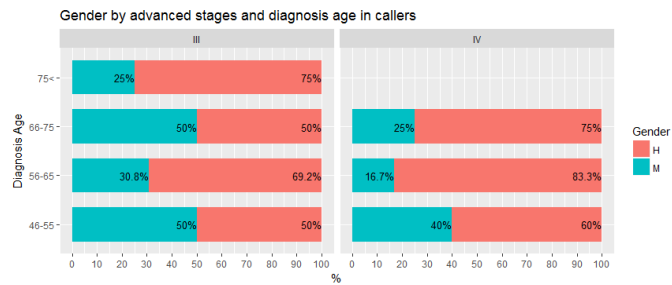

Fig. 11.  Distribution of gender by advanced stages and diagnosis age in callers.

As a final step, a spatial visualization was carried out, aimed at describing the relation of the localization of patients and the use of the oncology call system. Previously, we asked for the total number of lung cancer patients and the places where they live.

We also got the population data of those places. Using this information, it was possible to calculate the ratio of callers / non-callers per city regions - see Figure 13 for a representation. Once again, this kind of information is useful from the hospital staff's perspective, as it allows to characterize and have a more complete overview of the people using their system.

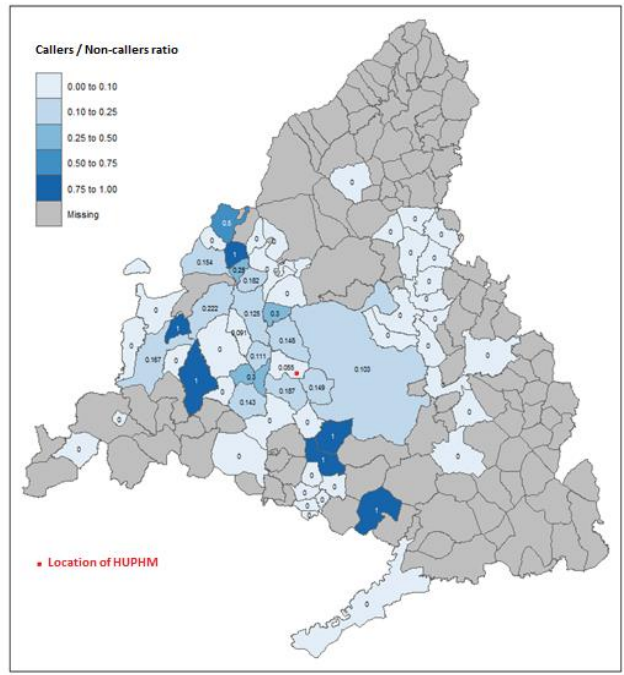

Fig. 12. Spatial representation of the ratio of callers / non-callers.

## 4. Analysis and results

The main objective of the study was the achievement of a better understanding of the patients who use the call service, through the application of the C-liKES system. The fact of having been able to extend the study and to carry out the analysis proves the usefulness of the system in this use case, and confirms its potential to satisfy needs of this nature in similar scenarios.

Beyond the results here reported, which just represent a first approach to a much more complex problem, the development of the descriptive analysis has allowed us to define and raise hypotheses based on the relationships and knowledge extracted from the data. While these cannot be tested in a reliable manner with the presently available data, some of them are discussed below, for potentially being the starting point of future studies:

Patients in advanced stages are more likely to use the call service, as the distribution according to the disease stage differs considerably from that of the whole population – as observed by physicians.

The median age at diagnosis of lung cancer in the studied patients is about 60 years, while the median age of lung cancer diagnostic is 72, according to the most recent statistics collected between 2008 and 2012 [11]. This suggests that the population of lung cancer patients who use the call service is younger than the entire population of lung cancer patients.

More impaired patients, as measured by the ECOG scale, seem to be less prone to use the system. While expected, this opens the door towards a campaign fostering caring people to use the system, when the patient him/herself cannot do that.

These raised hypotheses could later be validated through statistical studies, further aimed at demonstrating cause-effect relationships between the variables.

## Acknowledgements

This paper is supported by European Union's Horizon 2020 research and innovation programme under grant agreement No. 727658, project IASIS (Integration and analysis of heterogeneous big data for precision medicine and suggested treatments for different types of patients).

## Conflict of interest

The author(s) declare(s) that there is no conflict of interest regarding the publication of this paper.